\documentclass[twocolumn,trackchanges]{aastex62}
\usepackage{amssymb}
\usepackage{graphicx}
\usepackage{grffile}
\usepackage{epsfig}
\usepackage{epstopdf}
\usepackage[T1]{fontenc}
\usepackage{apjfonts}
\usepackage{url}

\newcommand{\msun}{\ensuremath{\rm M_\odot}}

\newcommand{\msunyr}{\ensuremath{\rm M_{\odot}\,{\rm yr}^{-1}}}

\newcommand{\Hb}{\ensuremath{\rm H\beta}}

\newcommand{\lya}{\ensuremath{\rm Ly\alpha}}
\newcommand{\wlya}{$W_{\rm Ly\alpha}$}

\newcommand{\kms}{km\,s\ensuremath{^{-1}}}

\newcommand{\OII}{[\ion{O}{2}]}
\newcommand{\OIII}{[\ion{O}{3}]}
\newcommand{\OIIIuv}{\ion{O}{3}]}
\newcommand{\CIII}{\ion{C}{3}]}
\newcommand{\CIV}{\ion{C}{4}}
\newcommand{\HeII}{\ion{He}{2}}
\newcommand{\HII}{\ion{H}{2}}
\newcommand{\HI}{\ion{H}{1}}

\newcommand{\expnt}[2]{\ensuremath{#1 \times 10^{#2}}}   
\newcommand{\fluxunits}{\ensuremath{\mathrm{erg}\,\mathrm{s}^{-1}\,\mathrm{cm}^{-2}}}
\newcommand{\flamunits}{erg s$^{-1}$ cm$^{-2}$ \AA$^{-1}$}
\newcommand{\sbunits}{\ensuremath{\mathrm{erg}\,\mathrm{s}^{-1}\,\mathrm{cm}^{-2}\,\mathrm{arcsec}^{-2}}}
\newcommand{\sbflamunits}{erg s$^{-1}$ cm$^{-2}$ \AA$^{-1}$ arcsec$^{-2}$}

\newcommand{\sls}{SL2S J0217}

\shorttitle{Sub-kiloparsec Imaging of \lya\ Emission in a Lensed Galaxy}
\shortauthors{Erb et al.}

\turnoffeditone

\begin{document}

\title{\large \textbf{Sub-kiloparsec Imaging of Ly\boldmath{$\alpha$}\unboldmath\ Emission in a Low Mass,\\ Highly Ionized, Gravitationally Lensed Galaxy at \textit{z}~\boldmath{$= 1.84$}\unboldmath\footnote{Based on observations with the NASA/ESA Hubble Space Telescope  at the Space Telescope Science Institute, which is operated by the Association of Universities for Research in Astronomy, Incorporated, under NASA contract NAS5-26555. Support for program number 14632 was provided through a grant from the STScI under NASA contract NAS5-26555.}}}

\correspondingauthor{Dawn K. Erb}
\email{erbd@uwm.edu}

\author[0000-0001-9714-2758]{Dawn K. Erb}
\affil{The Leonard E.\ Parker Center for Gravitation, Cosmology and Astrophysics, Department of Physics, University of Wisconsin-Milwaukee, 3135 N Maryland Avenue, Milwaukee, WI 53211, USA}
\affil{Department of Astronomy, Oskar Klein Centre, Stockholm University, AlbaNova University Centre, SE-106 91 Stockholm, Sweden}

\author[0000-0002-4153-053X]{Danielle A. Berg}
\affiliation{Department of Astronomy, The Ohio State University, 140 W. 18th Avenue, Columbus, OH 43202}

\author{Matthew W. Auger}
\affil{Institute of Astronomy, University of Cambridge, Madingley Road, Cambridge, CB3 0HA, UK}

\author[0000-0001-6295-2881]{David L. Kaplan} 
\affil{The Leonard E.\ Parker Center for Gravitation, Cosmology and Astrophysics, Department of Physics, University of Wisconsin-Milwaukee, 3135 N Maryland Avenue, Milwaukee, WI 53211, USA}
\affil{Department of Astronomy, Oskar Klein Centre, Stockholm University, AlbaNova University Centre, SE-106 91 Stockholm, Sweden}

\author[0000-0003-2680-005X]{Gabriel Brammer}
\affil{The Cosmic Dawn Center, University of Copenhagen, Vibenshuset, Lyngbyvej 2, 2100 Copenhagen, Denmark}

\author[0000-0002-5139-4359]{Max Pettini}
\affil{Institute of Astronomy, University of Cambridge, Madingley Road, Cambridge, CB3 0HA, UK}

\begin{abstract}
Low mass, low metallicity galaxies at low to moderate ($z\lesssim3$) redshifts offer the best opportunity for detailed examination of the interplay between massive stars, ionizing radiation and gas in sources similar to those that likely reionized the universe. We present new narrowband \textit{Hubble Space Telescope} observations of \lya\ emission and the adjacent ultraviolet (UV) continuum in the low mass ($M_{\star} = 2 \times 10^8$ \msun), low metallicity ($Z\sim1/20$ Z$_{\odot}$) and highly ionized gravitationally lensed galaxy \object[SL2S J021737-051329 source]{SL2S J02176$-$0513} at $z=1.844$. The galaxy has strong \lya\ emission with photometric equivalent width $W^{\rm phot}_{\lya} = 218 \pm 12$ \AA, at odds with the \lya\ escape fraction of 10\%. However, the spectroscopic \lya\ profile suggests the presence of broad absorption underlying the emission, and the total equivalent width is consistent with the escape fraction once this underlying absorption is included. The \lya\ emission is more spatially extended than the UV continuum, and the 0\farcs14 spatial resolution of \textit{HST} coupled with the magnification of gravitational lensing enables us to examine the distribution of \lya\ and the UV continuum on sub-kiloparsec scales. We find that the peaks of the \lya\ emission and the UV continuum are offset by 650 pc, and there is no \lya\ emission arising from the region with the strongest UV light. Our combined spectroscopic and imaging data imply a significant range in neutral hydrogen column density across the object. These observations offer indirect support for a model in which ionizing radiation escapes from galaxies through channels with low column density of neutral gas. 
\end{abstract}


\section{Introduction}
\label{sec:intro}
Roughly 500 million years after the Big Bang, at a time corresponding to the emergence of the first sources of ionizing radiation, the neutral hydrogen gas pervading the universe became ionized (e.g., \citealt{planck18}). The identity of these sources of ionizing radiation remains uncertain, but it is likely that faint, low mass galaxies contributed a significant fraction of the required numbers of energetic photons \citep{kf12,refd15}. Thus low mass, low metallicity galaxies are a population of cosmological importance, and considerable effort has been expended in recent years on the attempt to understand the intrinsic ionizing spectra of such galaxies, as well as how or if this radiation escapes into the intergalactic medium (IGM).

Massive stars in low metallicity galaxies produce significant amounts of Lyman continuum radiation (LyC; photons with energy $E>13.6$ eV), but the likelihood of this radiation escaping from the galaxies depends on the amount and distribution of neutral hydrogen gas. Our ability to observe this ionizing radiation further depends on the opacity of the IGM along the line of sight, with the result that the Lyman continuum cannot be directly detected during the epoch of reionization. It is therefore essential to understand the physics of LyC escape in galaxies at lower redshifts, where this emission can be detected and measured.

Low mass, low metallicity, highly ionized galaxies in the local universe now account for most of the objects with individual detections of LyC radiation \citep{lbh+13,lhlo16,ios+16,ist+16,isw+18,iws+18}, and an understanding of the neutral gas properties of such objects is therefore paramount. In the absence of direct measurements of \HI, which are prohibitively difficult outside of the very nearby universe and even then may result in upper limits \citep{pho+17,mjo+19}, perhaps the best way to constrain the properties of neutral hydrogen is through the observation of the resonantly scattered \lya\ emission line.  

The expanding outflows of highly star-forming galaxies produce redshifted \lya\ emission lines, as the photons that most easily escape the galaxy are backscattered from the far side of the outflow (e.g., \citealt{pss+01,ssp+03,e15,g17}). As the opacity to \lya\ photons decreases a secondary blueshifted peak of emission may become increasingly prominent, and the separation between the blue and red peaks decreases as photons require fewer scatterings to escape the gas (e.g., \citealt{vsm06,d14}).  Thus observations of highly ionized local emission-line galaxies (e.g., the ``Green Peas") usually show strong, often double-peaked \lya\ emission \citep{hsme15,josd17,ymg+17}. 

We further expect the escape of \lya\ and LyC emission to be coupled, since both depend on the distribution of neutral gas \citep{vosh15}. This expectation is supported by the narrow \lya\ peak separation of known LyC-emitting galaxies in the local universe \citep{vos+17}, and by the increase in ionizing flux density relative to the non-ionizing continuum with increasing \lya\ equivalent width observed in galaxies at $z\sim3$ \citep{sbs+18}.

 A complete examination of the \lya\ profiles reveals a complex picture, however; some extremely low metallicity galaxies have damped \lya\ absorption (e.g., \citealt{jah+14}), while some of the sources with the highest degree of ionization in their \HII\ regions show a blend of strong emission and underlying absorption \citep{mjo+19}. It is clear that \HI\ gas with a wide range of column densities may exist in these extreme galaxies, and geometrical effects are a strong factor in determining ultimate \lya\ escape.

Because \lya\ photons are scattered spatially as well as spectrally, images of \lya\ emission provide constraints on the extent of neutral hydrogen gas.  \textit{Hubble Space Telescope (HST)} observations of small samples of local galaxies reveal extended \lya\ emission \citep{hod+14}, and \lya\ halos are now known to be ubiquitous around galaxies at high redshifts; such halos are detected in stacked, narrowband \lya\ images \citep{sbs+11,myh+12,mon+14,mon+16} and in observations of individual galaxies with sensitive new integral field spectrographs \citep{wbb+16,lbw+17,esy18}. The most powerful constraints on \HI\ from \lya\ therefore come from the combination of both spatial and spectroscopic observations.

In this paper we use both spectroscopic measurements of the \lya\ profile and the sub-kpc spatial resolution afforded by the combination of \textit{HST} and gravitational lensing to compare the properties of \lya, the UV continuum, and collisionally excited emission lines in the $z=1.844$ gravitationally lensed arc \object[SL2S J021737-051329 source]{SL2S J02176$-$0513} ($\alpha=$~02:17:37.237, $\delta = -$05:13:29.78, J2000; hereafter \sls). \sls\ was discovered by the Strong Lensing Legacy Survey \citep{tgl+09,cfc+11}, and has been observed extensively with \textit{HST} due to its location in the UKIDSS/UDS field, which has been covered by both the multi-wavelength CANDELS imaging survey \citep{candels-g,candels-k} and the 3D-HST spectroscopic grism survey \citep{3dhst2016}.

\sls\ has been previously studied by \citet{bsl+12} and \citet{bea+18}. The system consists of a low mass ($M_{\star} = 2 \times 10^8$ \msun), strongly star-forming ($\mbox{SFR} = 23$ \msunyr), low metallicity ($Z\sim1/20$ Z$_{\odot}$), and highly ionized galaxy at $z=1.8444$, magnified by a factor of $\sim17$ by a massive foreground galaxy at $z = 0.6459$. Many of the properties of the lensed arc are extreme:\ the high SFR and low stellar mass result in an exceptionally high specific star formation rate $\mbox{sSFR}=120$ Gyr$^{-1}$, while the 3D-HST grism spectrum reveals \OIII\ $\lambda\lambda$4959, 5007 emission with rest-frame equivalent width $W_0 = 2095$ \AA\ and line ratios similar to those found in very low metallicity starburst galaxies in the local universe \citep{bsl+12}. 

\begin{figure*}[htbp]
\centerline{\epsfig{angle=00,width=\hsize,file=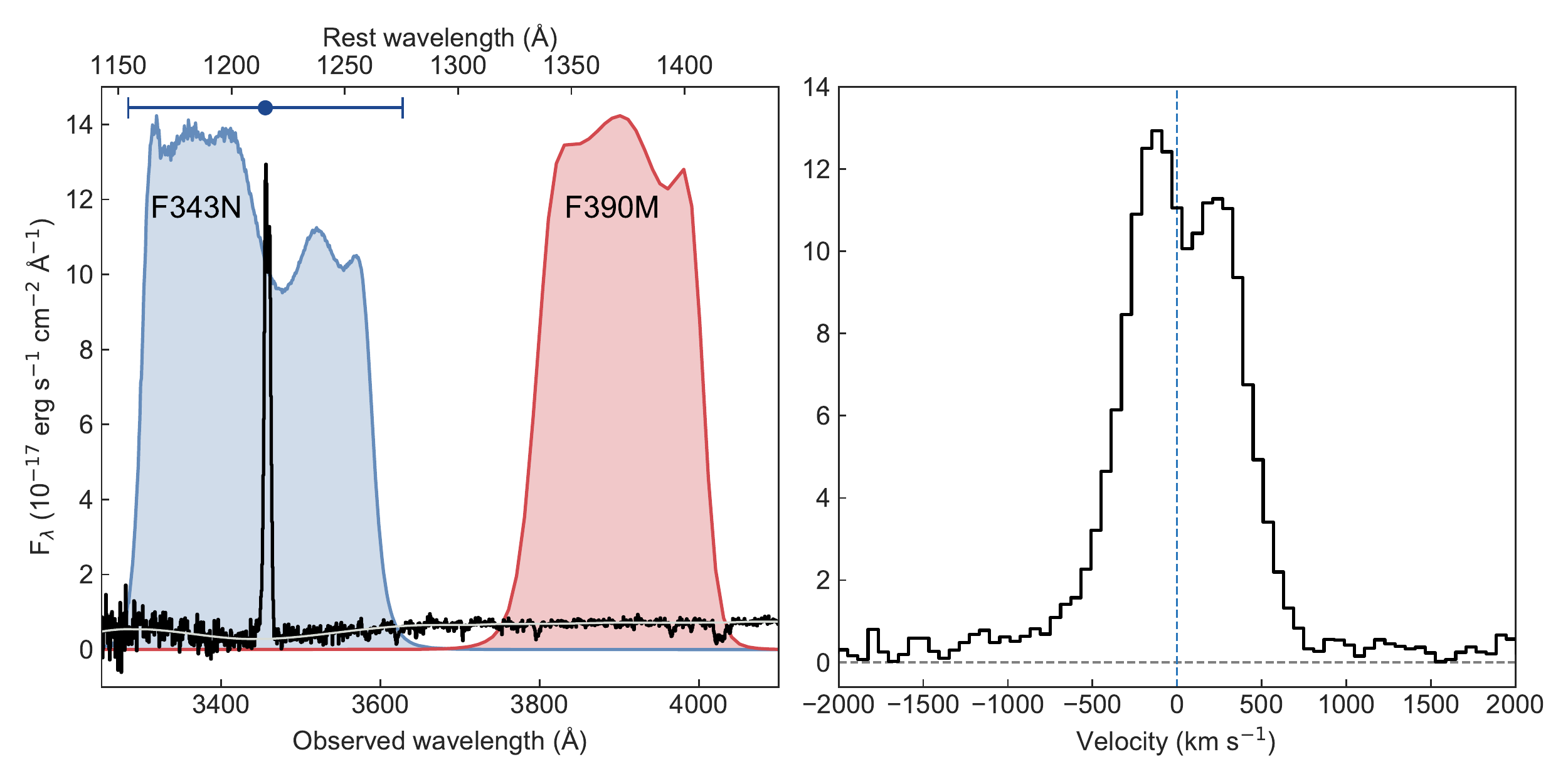}}
\caption{{\it Left:} The spectrum of \sls\ with the F343N and F390M bandpasses overlaid. The thin light grey line indicates the continuum fit used to separate the line and continuum flux in the F343N filter, and the blue point with horizontal error bar shows the spectrum-weighted average wavelength $\bar \lambda$ and the rectangular-equivalent filter width $\Delta \lambda$; see Section \ref{sec:lyaflux}. {\it Right:} The double-peaked \lya\ profile in velocity space.}
\label{fig:filters_spectrum}
\end{figure*}

\citet{bea+18} have recently presented a detailed analysis of the rest-frame UV spectrum of \sls\ from the Low Resolution Imaging Spectrograph (LRIS) on the Keck I telescope. This spectrum is dominated by strong emission from nebular \CIV\ $\lambda\lambda$1548,1550, \HeII\ $\lambda$1640, \OIIIuv\ $\lambda\lambda$1661,1666, and \CIII\ $\lambda\lambda$1907,1909, indicating a very hard ionizing spectrum.  \sls\ is thus part of a growing sample of low metallicity galaxies at low (e.g., \citealt{jo14,josd17,bsh+16,iws+18,ssc+19,bce+19}), intermediate ($z\sim2$--$3$; \citealt{eps+10,stark+14,eps+16,vdc+16,afp+17}) and high ($z\gtrsim6$; \citealt{swc+15,src+15,sec+17,mzs+18}) redshifts characterized by extreme ionization and high equivalent width nebular line emission.
\sls\ also shows strong, double-peaked \lya\ emission with a slightly stronger blue peak and interstellar absorption lines consistent with the systemic velocity, indicating that in spite of the strong star formation there is no evidence for outflowing gas along the line of sight \citep{bea+18}. 

The paper is arranged as follows. We describe our observations and the reduction of the new {\it HST} data in Section \ref{sec:data}.  In Section \ref{sec:strength} we assess the strength of the \lya\ emission, measuring the total photometric \lya\ flux in Section \ref{sec:lyaflux} and examining the \lya\ escape fraction and equivalent width in Section \ref{sec:wlya}. We quantify the spatial distribution of both \lya\ and the UV continuum in Section \ref{sec:spatial}, focusing on the image plane in Section \ref{sec:image} and the source plane in Section \ref{sec:source}. In Section \ref{sec:discuss} we summarize our results and discuss their implications. We assume that $H_0=70$ \kms\ Mpc$^{-1}$, $\Omega_{\rm m}=0.3$, and $\Omega_{\Lambda}=0.7$. With these values of the cosmological parameters, 1\arcsec\ at the redshift of \sls\ corresponds to 8.4 proper kpc. 

\begin{figure*}[htbp]
\centerline{\epsfig{angle=00,width=\hsize,file=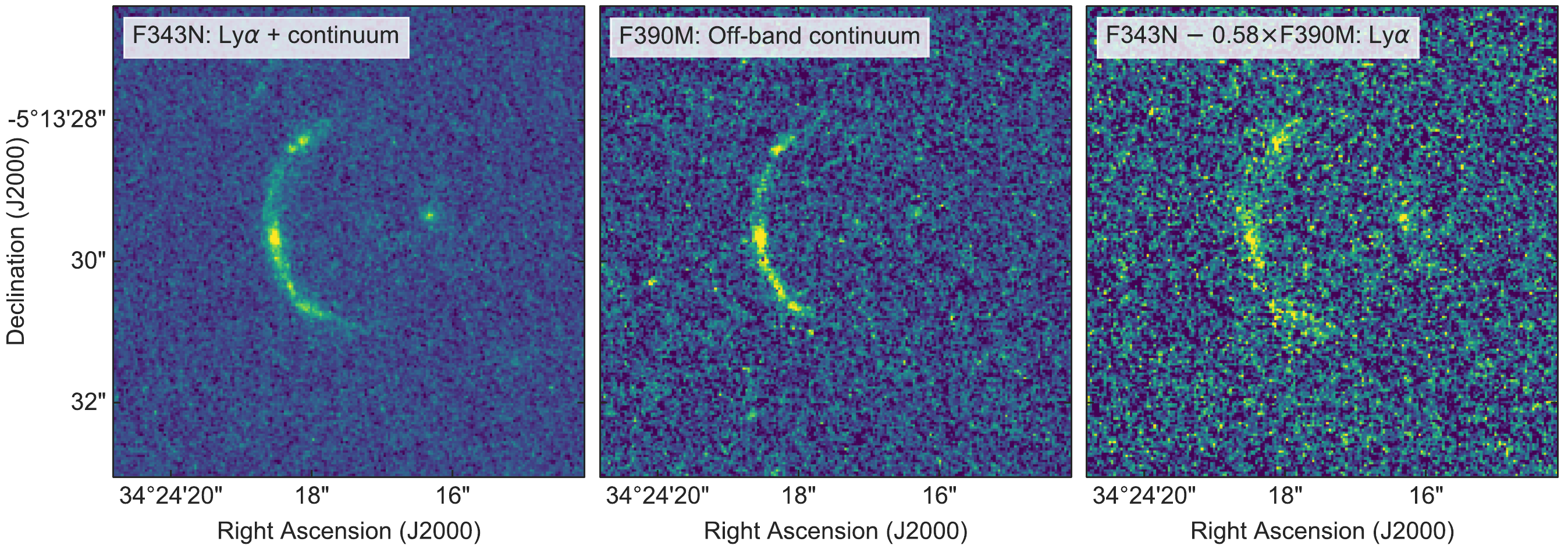}}
\caption{New {\it HST} imaging of \sls. The object to the right of the arc is the counterimage of the background source. The massive red $z=0.65$ lensing galaxy is invisible at the rest-frame near-UV wavelengths covered by the two filters. {\it Left:} The F343N image, including the \lya\ emission line. {\it Center:} The F390M image, covering the adjacent UV continuum. {\it Right:} The continuum-subtracted \lya\ image, with the continuum level determined as described in the text.}
\label{fig:3panel_images}
\end{figure*}

\section{Observations and Data Reduction}
\label{sec:data}
We have obtained new images of \sls\ with the UVIS channel of the Wide Field Camera 3 (WFC3) on the \textit{Hubble Space Telescope}, using the F343N and F390M filters (see Figure \ref{fig:filters_spectrum}) to observe \lya\ and the adjacent UV continuum respectively (program ID GO-14632, PI Erb). Observations in the F343N filter consist of six orbits with two dithered exposures per orbit, for a total F343N integration time of 16,758 s. In the F390M filter we obtained one orbit with two dithered exposures, for a total F390M integration time of 2738 s.

We combine the raw exposures in each filter using \texttt{AstroDrizzle} \citep{2012AAS...22013515H,2015ASPC..495..281A} version 2.1.21.  We first astrometrically align the images from each independent visit, and then combine the exposures setting drizzle $\rm pixfrac=1.0$ and using a final pixel scale of $0\farcs033$. We use inverse variance weighting and do not subtract the sky background from the raw images in order to preserve the noise statistics for later analysis. We instead subtract the background from the drizzled images as a final step, using background images constructed with \texttt{SExtractor} \citep{ba96}. Before making photometric measurements we use the \citet{sf11} dust map to correct the F343N and F390M images for foreground Galactic extinction, finding $A_V = 0.0601$ along the line of sight to \sls. Using the \citet{ccm89} extinction law, the resulting corrections are 0.096 and 0.090 mag at the effective wavelengths of the F343N and F390M filters respectively. 

The fully reduced images have units of counts per second, which can be converted to flux density using the flux density of a source with $1\,\mathrm{count}\,\mathrm{s}^{-1}$ in the appropriate filter.\footnote{These flux densities are $\expnt{2.53}{-18}$ \flamunits\ for F343N and  $\expnt{2.51}{-18}$ \flamunits\ for F390M, also given as \texttt{PHOTFLAM} in the image headers.} The $1\sigma$ limiting specific surface brightnesses of the images are $1.4\times10^{-18}$ \sbflamunits\ and $3.3\times10^{-18}$ \sbflamunits\ in the F343N and F390M filters respectively.

We also make use of broadband imaging in the ACS F606W, ACS F814W and WFC3 F160W filters from CANDELS \citep{candels-g,candels-k}, and WFC3 grism observations from the 3D-HST survey \citep{3dhst2016} previously described by \citet{bsl+12}. Finally, we use the rest-frame UV Keck/LRIS spectrum of \sls\ previously analyzed by \citet{bea+18}. This spectrum is shown in Figure \ref{fig:filters_spectrum}.

\section{The Strength of \lya\ Emission}
\label{sec:strength}

We use the new narrowband imaging of \sls\ to measure the \lya\ flux and equivalent width in the F343N filter, informed by the spectrum shown in Figure \ref{fig:filters_spectrum}. Briefly, we create a continuum-subtracted \lya\ image and define a region encompassing the arc, and then measure the total \lya\ flux in two different ways for verification. We scale a synthetic spectrum integrated over the filter bandpass until it matches the counts in the image, and additionally determine the effective filter width needed to calculate the total \lya\ flux. Finally we measure the continuum level of the spectrum in order to compute the equivalent width. We describe each of these steps in more detail below, and all measured and derived quantities are reported in Table \ref{tab:measurements}.

\subsection{Photometric measurement of Ly$\alpha$ emission}
\label{sec:lyaflux}
In order to measure the total \lya\ flux captured by the F343N filter, we first use the spectrum to determine the fraction of the flux in the filter due to the continuum. We mask the significant emission and absorption lines and fit the resulting masked spectrum with a spline; this fit is shown by the grey line in the left panel of Figure~\ref{fig:filters_spectrum}. We then use this spline to interpolate the spectrum over the Ly$\alpha$ emission line to form a line-free continuum spectrum.

We determine the continuum count-rate in each filter $\mathrm{CR}_{\mathrm{F343N},\mathrm{line-free}}$ and $\mathrm{CR}_{\mathrm{F390M},\mathrm{line-free}}$ by integrating this line-free spectrum over the F343N and F390M bandpasses taken from \texttt{pysynphot} and using the {\it HST} collecting area.  The ratio of these two count-rates then gives the relative strength of the continuum in the F343N and F390M filters. We find a continuum correction factor
\begin{equation}
\mathrm{CCF} = \frac{\mathrm{CR}_{\mathrm{F343N},\mathrm{line-free}}}{\mathrm{CR}_{\mathrm{F390M},\mathrm{line-free}}} = 0.58.
\label{eq:ccf}
\end{equation}
We then make an F343N continuum image by scaling the F390M image by this correction factor, and finally create the line-only \lya\ image by subtracting this continuum image from the F343N image: $\mathrm{F343N}(\lya) = \mathrm{F343N} - \mathrm{CCF} \times \mathrm{F390M}$. This image is shown in the rightmost panel of Figure \ref{fig:3panel_images}. We note that the continuum in F343N is significantly lower than in F390M (i.e.\ CCF~$<1$), \edit1{and that we have assumed that the continuum correction factor is uniform across the image}; we discuss \edit1{these issues further in Sections \ref{sec:wlya} and \ref{sec:discuss} respectively.}

The next step is to identify the pixels belonging to the arc. We define an initial region using pixels with S/N~$>2$ in the F343N image smoothed with a FWHM~$=4$ pixel Gaussian filter, and then apply this region to the unsmoothed line-only \lya\ image and increase the size of the region until the counts in the \lya\ image are maximized\footnote{Note that the background of the \lya\ image is consistent with zero.} (this is also the point at which the S/N of the enclosed \lya\ emission begins to drop). Applying this mask to the F343N and F390M images, we measure the AB magnitudes listed in Table \ref{tab:measurements}.

The total \lya\ flux in the F343N filter can be determined by constructing a synthetic spectrum with an emission line with the observed width and wavelength of \lya, integrating this spectrum over the F343N bandpass, and scaling the emission line until the count-rate obtained from the synthetic spectrum matches the count-rate in the masked, continuum-subtracted \lya\ image. This total flux is $F^{\rm phot}_{\lya} = \expnt{(1.70\pm0.09)}{-15}\,\fluxunits$.

We can also determine the flux in the \lya\ image by multiplying the flux density in the line-only image by the appropriate filter width. We use the bandpass information from \texttt{pysynphot} to calculate the effective rectangular filter width  
\begin{equation}
\Delta \lambda = \frac{\int R_\lambda(\lambda)\,d\lambda}{R_\lambda(\bar \lambda_{\mathrm{F343N}})},
\label{eq:filtwidth}
\end{equation}
where $R_\lambda(\lambda)$ is the filter response evaluated at $\lambda$ and $\bar \lambda_{\mathrm{F343N}}$ is the average wavelength of the spectrum weighted by the filter transmission.  Note that this differs from the \texttt{pysynphot} version of this definition,\footnote{See \url{https://pysynphot.readthedocs.io/en/latest/properties.html\#}.} which uses the peak response of the filter.  That definition would be appropriate for a smooth, continuum-dominated spectrum, but as is visible in Figure~\ref{fig:filters_spectrum} the peak response of the filter does not coincide with the wavelength of the Ly$\alpha$ emission line.

The filter-weighted average wavelength is found by integrating the continuum-subtracted spectrum over the filter,
\begin{equation}
\bar \lambda_{\mathrm{F343N}}= \frac{\int\lambda R_\lambda(\lambda)F_{\lambda}^{\mathrm{line-only}}\,d\lambda}{\int R_\lambda(\lambda) F_{\lambda}^{\mathrm{line-only}}\,d\lambda},
\end{equation}
for which we find $\bar \lambda_{\mathrm{F343N}}=3455.6\,$\AA. With this average wavelength the effective rectangular filter width is $\Delta \lambda_{\mathrm{F343N}}=344.6\,$\AA, compared to $249.4\,$\AA\ based on the standard definition.  The product of the line-only image and this rectangular filter width is the final \lya\ image; this image has $1\sigma$ limiting surface brightness $8.3\times10^{-16}$ \sbunits. 

To calculate the \lya\ flux in the F343N image we apply the spatial mask to the line-only image in order to isolate the Ly$\alpha$ emission.  We compute the count-rate by spatially summing the Ly$\alpha$ emission in the masked line-only image, and convert this into a flux density as described in Section \ref{sec:data}. The \lya\ flux is then the product of this flux density and the effective rectangular filter width $\Delta \lambda$ defined in Equation \ref{eq:filtwidth}. The result is $F^{\rm phot}_{\lya} = \expnt{(1.69\pm0.09)}{-15}\,\fluxunits$, in agreement with the flux obtained from the synthetic emission line above.

Finally, we measure the \lya\ flux from the spectrum in order to assess the impact of spectroscopic slit losses. We first adjust the flux calibration of the spectrum to match the F390M continuum image by multiplying the spectrum by a factor of 0.96; this is the ratio of the count-rate in the F390M image and the count-rate in the synthetic F390M image created by integrating the spectrum over the filter. After this small correction, we measure the line flux in the  spectrum to be $F^{\rm spec}_{\lya} = \expnt{(1.1\pm0.3)}{-15}\,\fluxunits$, or $(66\pm16)$\% of the flux in the \lya\ image.  We use the final \lya\ image and the effective rectangular filter width for spatially-resolved surface brightness measurements in Section \ref{sec:spatial} below.

\begin{deluxetable}{l r}
\tablewidth{0pt}
\tabletypesize{\footnotesize}
\tablecaption{Photometric and Spectroscopic Measurements\label{tab:measurements}}
\startdata
& \\
F343N AB magnitude & $22.54 \pm 0.01$ \\
F390M AB magnitude & $22.59 \pm 0.07$ \\
Spectroscopic \lya\ flux (\fluxunits) & $(1.12 \pm 0.26) \times10^{-15}$ \\
Photometric \lya\ flux (\fluxunits) & $(1.69 \pm 0.09) \times10^{-15}$ \\
Spectroscopic emission-only \wlya\ (\AA) & $144 \pm 34$\\
Photometric emission-only \wlya\ (\AA) & $218 \pm 12$\\ 
Total emission+absorption \wlya\  (\AA) & $51 \pm 23$\\
\lya\ escape fraction & $0.097 \pm 0.006$ \\
\enddata
\label{tab:measurements}
\end{deluxetable}

\begin{figure*}[htbp]
\centerline{\epsfig{angle=00,width=\hsize,file=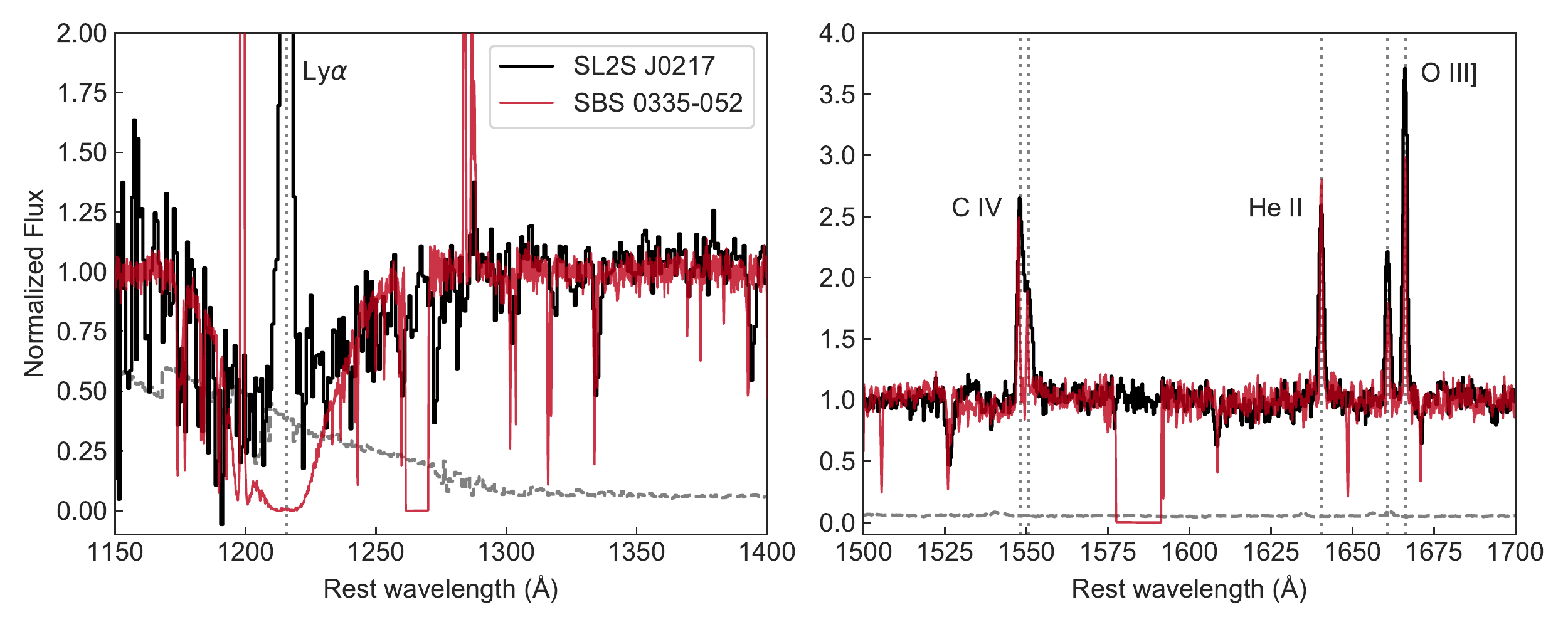}}
\caption{The Keck LRIS spectrum of \sls\ (black) compared with the \textit{HST} COS spectrum of the local low metallicity galaxy SBS 0335$-$052 (red). Both spectra are normalized, and the grey dashed lines in both panels show the \sls\ $1\sigma$ error spectrum. {\it Left:} The \lya\ profiles. The \sls\ spectrum has been rebinned by a factor of three to increase the signal-to-noise. The spectrum of SBS 0335$-$052 is from \textit{HST} program 11579 and has been presented by \citet{jah+14}. The strong emission features in the spectrum of SBS 0335$-$052 are geocoronal \lya\ and \ion{O}{1} emission. {\it Right:} The high ionization emission lines \CIV\ $\lambda\lambda$1549, 1551, \HeII\ $\lambda1640$, and \OIIIuv\ $\lambda\lambda$1661, 1666. The spectrum of SBS 0335$-$052 is drawn from the \textit{HST} archive (program 13788, PI Wofford). Rest wavelengths of the lines are marked with vertical dotted grey lines in both panels. }
\label{fig:sbs_compare}
\end{figure*}

\subsection{The Ly$\alpha$ escape fraction and equivalent width}
\label{sec:wlya}

We estimate the \lya\ escape fraction by predicting the expected \lya\ flux from the \Hb\ emission measured in the slitless WFC3 grism spectrum \citep{bsl+12}.  \edit1{Correcting for foreground Galactic extinction only} and assuming an intrinsic ratio \lya/\Hb~$=23.3$,\footnote{Appropriate to an electron density $n_e=100$ cm$^{-3}$; assuming $n_e=1000$ cm$^{-3}$ decreases the derived escape fraction by 10\%.} we find $f_{\rm esc}^{\lya}=0.097 \pm 0.006$, using the photometric \lya\ flux determined above. 

We compare this escape fraction with the \lya\ equivalent width, determined by dividing the \lya\ flux by the continuum flux density at the center of the line. Using the photometric and spectroscopic fluxes, the equivalent widths are $W^{\rm phot}_{\lya} = 218 \pm 12$ \AA\ and $W^{\rm spec}_{\lya} =144 \pm 34$ \AA\ respectively.

The \lya\ escape fraction and equivalent width are well known to be strongly correlated (e.g., \citealt{vos+17,hos+18,sm19}), and based on these correlations one would expect a galaxy with $W_{\lya} \approx 200$ \AA\ to have an escape fraction approaching 100\%, drastically different from the 10\% we measure for \sls. This discrepancy was noted previously by \citet{bea+18}, and we confirm here that it is still present when accounting for \lya\ flux missed by the spectroscopic slit. We note that additional low surface brightness \lya\ emission may still be present below our detection threshold, \edit1{if additional flux is scattered to large radii in the CGM}.

However, the emission-only equivalent width that we have measured above may not be a full representation of the \lya\ profile of \sls. As noted above, the continuum count-rate in the F343N filter predicted from the spectrum is only 58\% of the count-rate in the F390M filter, and examination of the spectrum and the continuum fit in Figure \ref{fig:filters_spectrum} shows a potential very broad absorption feature underlying the emission line. Broad \lya\ absorption is seen in other low metallicity, highly ionized galaxies:\ \citet{mjo+19} have recently presented examples of \lya\ emission with underlying absorption in extreme Green Pea galaxies, and in Figure \ref{fig:sbs_compare} we show the \lya\ profile of SBS 0335$-$052 \citep{jah+14}, one of the most metal-poor galaxies known. As discussed by \citet{bsl+12}, SBS 0335$-$052 is one of a very small number of local galaxies with rest-frame optical line ratios and spectral energy distribution (SED) similar to \sls.

As shown in the right panel of Figure \ref{fig:sbs_compare}, the high ionization UV emission lines of \sls\ and SBS 0335$-$052 are very similar. Both galaxies show narrow, nebular \HeII\ emission as well as emission from \CIV\ and \OIIIuv; the \HeII\ and \CIV\ emission in particular indicate a very hard ionizing spectrum \citep{bea+18}. In the left panel of the figure we compare the \lya\ profiles of the two objects. In spite of its high ionization SBS  0335$-$052 is a strong \lya\ \textit{absorber}, with a damped \lya\ profile indicating a column density $\log(N_{\rm HI}) = 21.7$ \citep{jah+14}. 

Although the \sls\ spectrum surrounding the \lya\ emission line is quite noisy, the general shape of the profile is similar to that of SBS  0335$-$052, suggesting that \sls\ may in fact be best characterized by a superposition of strong \lya\ emission and damped absorption.  Given the low S/N we do not attempt to determine the column density, but we do make a rough estimate of the total \lya\ equivalent width including this potential absorption in addition to the emission (see, e.g., \citealt{kse+10} for discussion of the complexity of the \lya\ profile and associated measurements of equivalent width). We define lower and upper limits of the absorption feature at 1170 \AA\ and 1292 \AA\ in the rest frame respectively; these are approximately the points at which the spectrum reaches the continuum on either side of the line. We then define the continuum surrounding \lya\ as the average value of the spectrum at these wavelength limits, subtract this average value, and integrate the spectrum between the wavelength limits to determine a net \lya\ flux. We multiply this flux by a factor of 1.5 to account for spectroscopic slit losses as discussed above, and divide the resulting value by the average continuum level to obtain an estimate of the total emission + absorption \lya\ equivalent width. The result is $W^{\rm tot}_{\lya} = 51 \pm 23$ \AA. 

\begin{figure*}[htbp]
\centerline{\epsfig{angle=00,width=\hsize,file=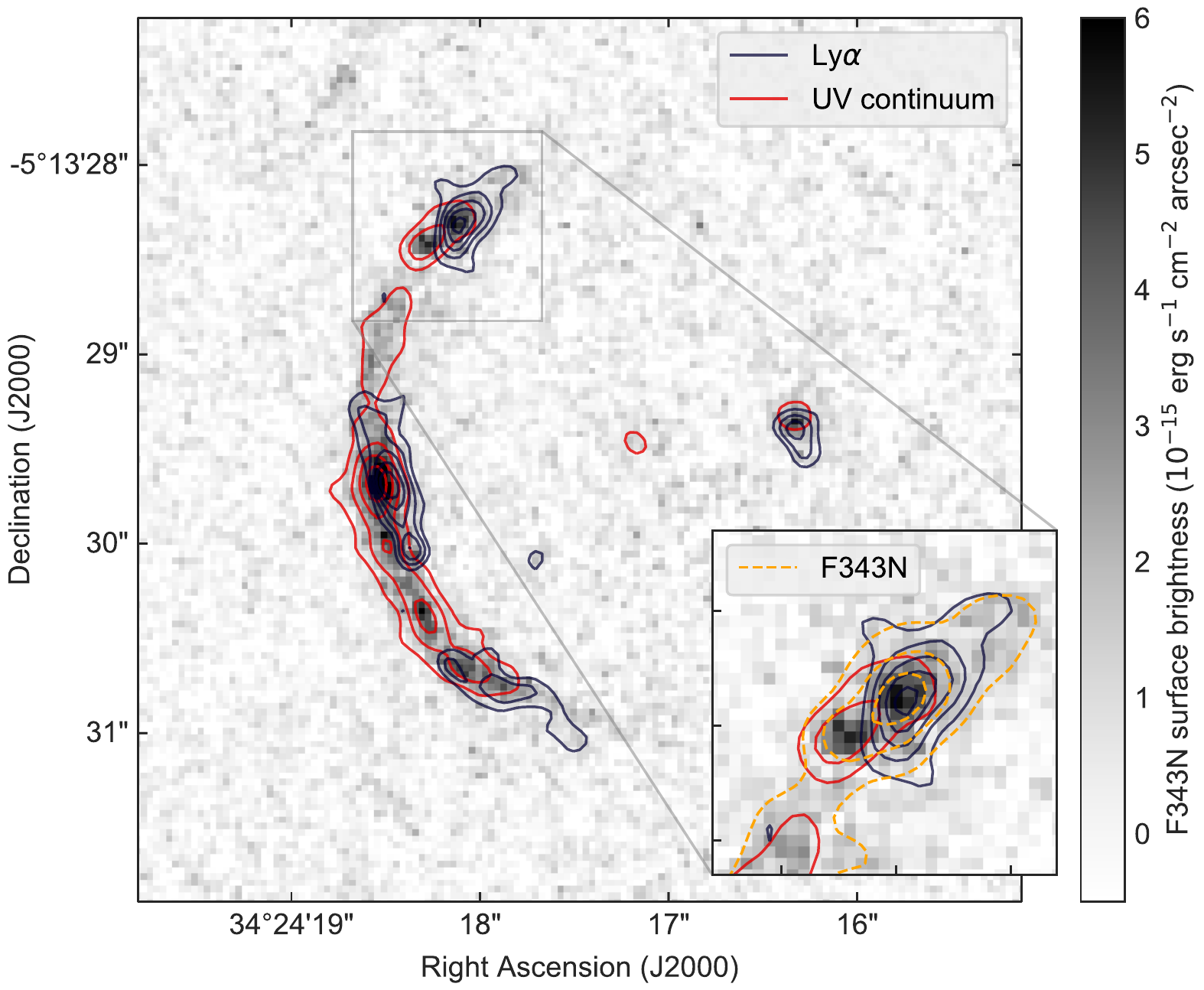}}
\caption{Contours of \lya\ (dark blue) and UV continuum (red) emission, shown on the \edit1{F343N} surface brightness image. Contours are calculated for the \lya\ and UV continuum images smoothed with a Gaussian filter with FWHM~$=4$ pixels (0\farcs13), and have levels (0.9,1.2, 1.5, 1.8, 2.1)$\times10^{-15}$ \sbunits\ for the \lya\ emission and (0.4, 0.8, 1.2, 1.6)$\times10^{-17}$ \sbflamunits\ for the continuum. \edit1{The inset at lower right shows the northern portion of the arc for which the \lya\ and continuum emission are most clearly separated, with additional dashed orange contours indicating the total emission in the F343N filter. F343N contour levels are (1, 2, 3)$\times10^{-15}$ \sbunits.}}
\label{fig:contours}
\end{figure*}

While this value is highly uncertain, it does indicate that including the underlying absorption can lead to a large reduction in equivalent width,  alleviating the discrepancy between the emission-only equivalent width and the escape fraction. Using the relationship between \lya\ equivalent width and escape fraction found by \citet{sm19}, $f_{\rm esc}^{\lya} = 0.0048 \times (W_{\lya}/\mbox{\AA}) \pm 0.05$, we would predict $f_{\rm esc}^{\lya} = 0.24 \pm 0.12$; this result differs from our measured value of $f_{\rm esc}^{\lya} = 0.097 \pm 0.006$ by $1.2\sigma$.  \edit1{We also note that we would have arrived at a similar total equivalent width by calculating $W_{\lya}$ from photometric information only; disregarding the spectrum and simply assuming that the continuum level is the same in the F343N and F390M filters results in an equivalent width of 41 \AA. As this method is closer to that employed for much of the \citet{sm19} sample, the improved agreement with their prediction is perhaps unsurprising. It is clear that the \lya\ equivalent width is highly sensitive to the method of measurement when complex profiles are involved.} We will return to the question of \lya\ absorption in Sections \ref{sec:source} and \ref{sec:discuss} below.

\begin{figure*}[htbp]
\centerline{\epsfig{angle=00,width=\hsize,file=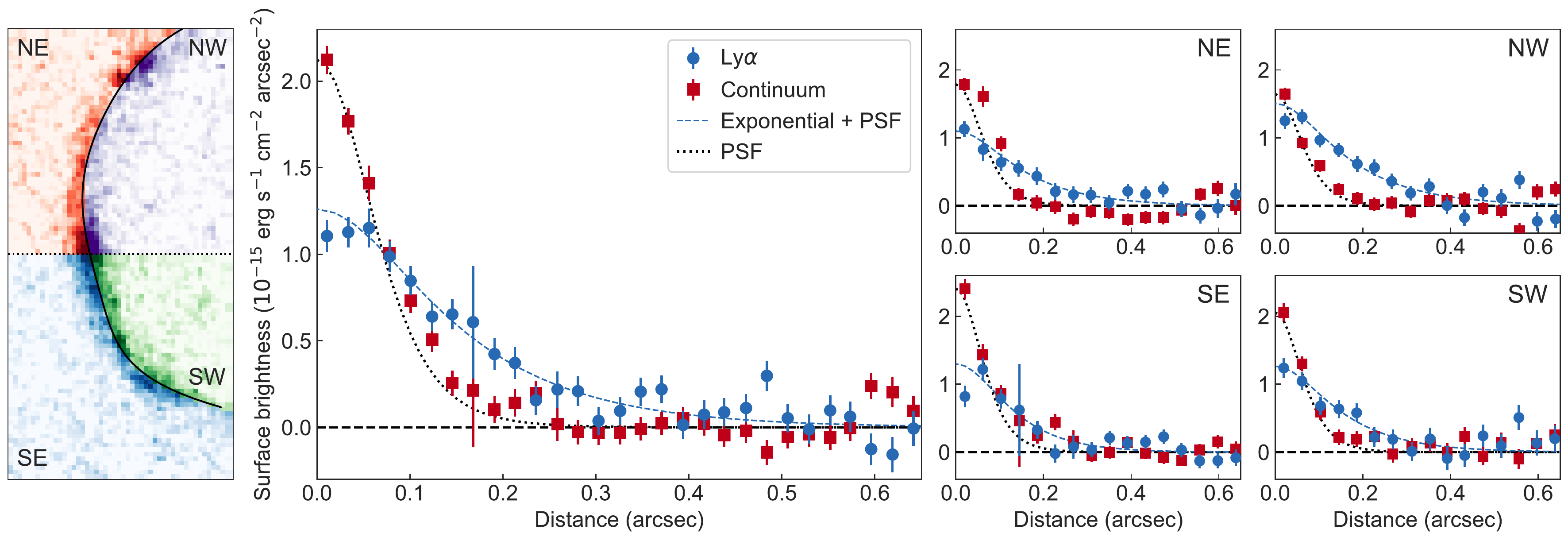}}
\caption{{\it Left:} Image of the arc with a solid black line showing the fit to the maximum flux along the arc. We further divide the arc into four quadrants using this ridgeline and the black dotted line at the vertical center of the arc. The four quadrants are plotted using different color scales and labeled in each corner of the image (red in the northeast at upper left, purple in the northwest at upper right, blue in the southeast at lower left, and green in the southwest at lower right). {\it Center:} Radial profiles of the \lya\ (blue circles) and UV continuum (red squares) surface brightness, measured perpendicular to the fit along the arc shown in the left panel. The PSF is shown by the black dotted line, and the blue dashed line shows the best fit to the \lya\ surface brightness profile using an exponential convolved with the PSF. {\it Right:} Radial profiles of the \lya\ and continuum emission surface brightness in each of the four quadrants of the arc separately (northeast, northwest, southeast and southwest, as labeled in the upper right corner of each panel and corresponding to the regions in the left panel).  Symbols are the same as in the center panel.}
\label{fig:radialprofiles}
\end{figure*}

\section{The Spatial Distribution of \lya\ Emission}
\label{sec:spatial}

The new {\it HST} imaging enables us to compare the relative spatial distributions of \lya\ emission and the continuum light from massive stars at $\sim1350$--1400 \AA. We first examine the extent of the \lya\ and continuum emission in the image plane, and then use the lensing model to reconstruct  images of the source in Section \ref{sec:source}.

\subsection{Ly$\alpha$ and the UV continuum in the image plane}
\label{sec:image}
A comparison of \lya\ and the UV continuum across the arc is shown in Figure \ref{fig:contours}, in which we draw contours of \lya\ and UV continuum surface brightness (from the continuum-subtracted \lya\ and F390M continuum images respectively) on the \edit1{F343N} image. As already suggested by the spectroscopic slit losses described above, the \lya\ emission is more extended than the continuum \edit1{at the ends of the arc}, and we also see evidence that the peaks of \lya\ and the continuum are not spatially coincident, particularly in the northern portion of the arc. \edit1{We zoom in on this portion of the arc in the inset panel at lower right in Figure \ref{fig:contours}, where we also show contours indicating the total emission in the F343N filter; these contours indicate that even before the continuum subtraction, the peaks of the F390M and F343N emission are spatially offset.}

In order to quantify the relative extents of \lya\ and continuum emission in the image plane, we construct radial profiles perpendicular to a line of maximum flux running along the arc.  We first use the F343N image to identify the brightest pixel in each row along the arc, and then fit a smooth polynomial to these points, weighting each point by its S/N.  This curve is shown in the left panel of Figure \ref{fig:radialprofiles}.

Next we determine the perpendicular distance of each pixel from this line, bin the pixels by this distance, and determine the mean surface brightness in each bin. We use bin sizes of 0\farcs02 for both the continuum-subtracted \lya\ and F390M continuum (rest-frame $\sim1350$--$1400$ \AA) images, and multiply both the \lya\ and continuum surface brightnesses by the appropriate effective rectangular filter widths determined as described in Section \ref{sec:lyaflux} above so that both surface brightnesses have units of \sbunits. The resulting radial profiles of \lya\ and UV continuum emission are shown in the middle panel of Figure \ref{fig:radialprofiles}.

These radial profiles of \lya\ and the continuum confirm the greater spatial extent of the \lya\ emission. We also compare the profiles with the point spread function, which we determine from the single bright star in the F343N and F390M images to be well-fit by a Moffat profile with FWHM~$=0\farcs14$ (black dotted line in Figure \ref{fig:radialprofiles}). The radial profile of the UV continuum (red squares) is barely distinguishable from this PSF, indicating that the continuum emission is largely unresolved in the direction perpendicular to the arc. In contrast, the \lya\ emission is not consistent with the PSF, and we find that the radial \lya\ profile is generally well-fit by an exponential with scale length $0\farcs12 \pm 0\farcs007$ convolved with the point spread function (dashed blue line). Thus although the \lya\ emission is more extended than the continuum, the relevant spatial scale in the direction perpendicular to the arc is still very small; if the lensing magnification is minimal in this perpendicular direction, this would correspond to a physical scale length of \edit1{$\sim1$ kpc}.

We next look for further spatial variations in the emission by dividing the arc into the four quadrants shown in the left panel of Figure \ref{fig:radialprofiles}, and constructing radial profiles as described above for the \lya\ and continuum emission in each quadrant separately. In this case we use larger bin sizes of 0\farcs04 in order to compensate for the lower total flux in each quadrant. The resulting profiles are shown in the four right-hand panels of Figure  \ref{fig:radialprofiles}.

The continuum emission is again largely consistent with the PSF in all quadrants, while the extent of \lya\ varies. We again fit an exponential convolved with the PSF to each radial profile, finding scale lengths of 0\farcs09--0\farcs12 in all quadrants except the northwest, for which we find a somewhat larger scale length of $0\farcs14 \pm 0\farcs01$. While these tests demonstrate that the \lya\ and UV continuum light do not arise from identical locations, further physical insight requires examining the images in the source plane.

\begin{figure*}[htbp]
\centerline{\epsfig{angle=00,width=\hsize,file=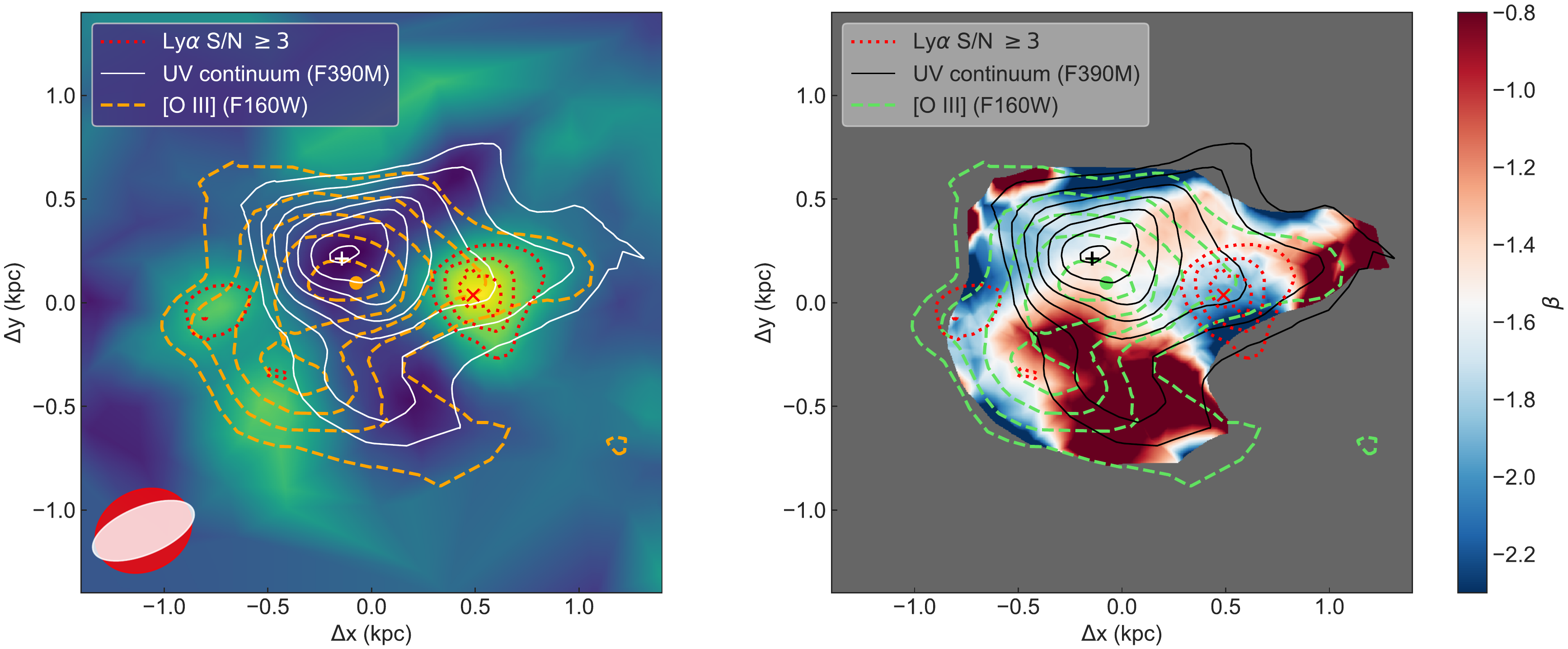}}
\caption{Source reconstructions from the lensing model. {\it Left:} Continuum-subtracted \lya\ surface brightness, with white contours outlining the UV continuum traced by the F390M filter and dashed orange contours indicating surface brightness in the F160W filter, which is dominated by \OIII\ $\lambda\lambda$4959, 5007 emission. UV continuum and F160W contour levels are (0.5, 1.0, 1.5, 2.0, 2.5, 3.0, 3.5)$\times10^{-17}$ and (0.1, 0.3, 0.5, 0.7, 0.9)$\times10^{-17}$ \sbflamunits\ respectively. Red dotted contours show regions of the \lya\ image with S/N~$>3$; contour levels are S/N~$=3$, 4 and 5. The red $\times$, white $+$ and orange dot indicate the \lya, UV continuum and F160W peaks respectively. \edit1{The red and white ellipses at lower left show the point spread function at the locations of the \lya\ and UV peaks respectively.} {\it Right:} The UV slope $\beta$ in the source plane, constructed from the F606W and F814W images which trace $\sim2000$--$2800$ \AA\ in the rest frame. We use only the regions of the images with surface brightness greater than $1.0 \times 10^{-18}$ \sbflamunits\ in both F606W and F814W, and extreme values of $\beta$ at the edge of the image are likely due to low S/N. Contours are the same as the left panel.}
\label{fig:source}
\end{figure*}

\subsection{Reconstructed source images}
\label{sec:source}
\citet{bea+18} present a reconstructed source image of SL2S J0217, using data from the CANDELS program and a new lensing model that assumes a lensing potential with an elliptical power-law mass distribution and an empirical external shear, while the source surface brightness is modeled as an irregular grid (see their Figure 1 and Section 3). This image reveals a compact, clumpy structure  $\lesssim1$ kpc in size, with significant variations in broadband color across the source. Here we apply the same lensing model to the new F343N and F390M data in order to assess the origin of the \lya\ and UV continuum emission in the source plane. 

We show the continuum-subtracted \lya\ surface brightness image in the source plane in the left panel of Figure \ref{fig:source}, with red dotted contours indicating the \lya\ S/N (note that because spatial resolution in the source plane depends on the S/N, the resolution of this image is lower than that of the image presented by \citealt{bea+18}). Additional contours show the UV continuum surface brightness in the F390M filter (solid white) and the surface brightness in the F160W filter (dashed orange) for comparison. As discussed by \citet{bsl+12}, the F160W filter is dominated by extremely strong \OIII$\lambda\lambda5007,4959$ emission with observed equivalent width 5690 \AA\ and flux $F_{\rm [OIII]} = 3.09\times10^{-15}$ \fluxunits. The AB magnitude F160W~$=20.90$ then implies that nebular emission accounts for $\gtrsim70$\% of the flux in the filter, and we use the orange contours as an estimate of the distribution of the non-resonant \OIII\ emission in the source plane. \edit1{The \lya\ and UV contours in Figure \ref{fig:source} approximately correspond to the regions traced by the dark blue and red contours in Figure \ref{fig:contours}.}

The reconstructed source image in Figure \ref{fig:source} shows that the peaks of the \OIII\ and UV continuum emission are largely spatially coincident, while the strongest \lya\ emission is offset to the west by 650 pc.\footnote{The spatial resolution in the source plane varies as a function of position, with FWHM~$\sim400$ pc at the locations of the \lya\ and UV peaks. The red and white ellipses at lower left in Figure \ref{fig:source} show the PSF at the \lya\ and UV peaks respectively, determined by simulating images with point sources at these locations and then finding the corresponding source-plane reconstructions. The \lya\ source is only marginally resolved, but it is clearly separable from the peak of the UV emission.} Furthermore we find that there is \textit{no} \lya\ emission arising from the regions of maximum UV and \OIII\ surface brightness; the continuum-subtracted \lya\ image is negative in the regions surrounding the UV and \OIII\ peaks. \edit1{This result is also suggested by the distribution of \lya\ and UV light in the image plane shown in Figure \ref{fig:contours}, in which we see that there is little or no \lya\ emission arising from the UV peak at the northern end of the arc.}

In the right panel of Figure \ref{fig:source} we construct a map of the UV slope $\beta$ in the source plane, using the F606W and F814W images which trace emission at $\sim2000$ \AA\ and $\sim2800$ \AA\ in the rest frame respectively. The variations in $\beta$ across the source are consistent with the color differences previously discussed by  \citet{bsl+12} and \citet{bea+18}, but we now see that the \lya\ emission arises primarily from a very blue portion of the galaxy with $\beta \approx -2$, while regions with the strongest UV and \OIII\ emission are redder with $\beta \approx -1.5$. \edit1{The UV slope of the full arc is intermediate between these values at $\beta \approx -1.7$, as reported by} \citet{bsl+12}. 

\edit1{Although we have followed previous spectroscopic work on \sls\ and assumed little to no dust, there is some uncertainty in the total extinction; in the low resolution grism spectrum H$\gamma$ is blended with [\ion{O}{3}] $\lambda$4363, making the reddening and metallicity determinations somewhat degenerate, while the extinction obtained from the SED fit is $A_V=0.09\pm0.15$ \citep{bsl+12}. The data thus allow for a small amount of extinction, as suggested by the redder regions of the image in both the source and image planes.}

The variations in the location of \lya\ emission compared to that of non-resonant nebular emission and the UV continuum presumably indicate small-scale relative differences between star-forming regions and their surrounding neutral hydrogen gas. We discuss these variations in Section \ref{sec:discuss} below, in which we bring together our spectroscopic and photometric results on the strength and spatial distribution of the \lya\ emission.

\section{Summary and discussion}
\label{sec:discuss}
We have presented new high resolution, narrowband \textit{HST} observations of \lya\ emission and the adjacent UV continuum in the low mass ($M_{\star} = 2 \times 10^8$ \msun), low metallicity ($Z\sim1/20$ Z$_{\odot}$) and highly ionized gravitationally lensed galaxy \object[SL2S J021737-051329 source]{SL2S J02176$-$0513} at $z=1.844$.  The rest-frame UV spectrum of the galaxy and the filters used for the new observations are shown in Figure \ref{fig:filters_spectrum}, and the \lya\ and continuum images are shown in Figure \ref{fig:3panel_images}. Our primary results are as follows:
\begin{enumerate}
\item{Comparison of the total flux in the \lya\ image with that in the spectrum indicates spectroscopic slit losses of $\sim33$\%, and the total \lya\ flux in combination with the continuum level underlying the emission line results in a total photometric equivalent width $W^{\rm phot}_{\lya} = 218 \pm 12$ \AA. This large equivalent width implies a \lya\ escape fraction of $\sim100$\%, at odds with the escape fraction $f_{\rm esc}^{\lya}=0.097 \pm 0.006$ we measure from the \lya/\Hb\ ratio.}
\item{The spectroscopic continuum surrounding the \lya\ emission line shows a broad trough similar in shape to the damped \lya\ profiles seen in some local low metallicity and highly ionized galaxies (see Figure \ref{fig:sbs_compare}), although higher continuum S/N is required in order to confirm this result. Including this absorption in the measurement of equivalent width results in an emission+absorption equivalent width $W^{\rm tot}_{\lya} = 51 \pm 23$ \AA, consistent within $\sim1\sigma$ with expectations from the \lya\ escape fraction given the uncertainties involved.}
\item{The \lya\ emission is more spatially extended than the UV continuum, with the largest differences seen in the northwest portion of the arc (Figures \ref{fig:contours} and \ref{fig:radialprofiles}). In the direction perpendicular to the arc, the \lya\ emission is well-fit with an exponential profile with scale length 0\farcs12 convolved with the point spread function, while the UV continuum is spatially unresolved.}
\item{Reconstructed source images of the continuum-subtracted \lya\ emission, the UV continuum, and the \OIII-dominated F160W image show that the peak of the \lya\ emission is offset from that of the UV continuum by 650 pc in the source plane, while the UV and \OIII\ peaks are largely spatially coincident. The \lya-only image reveals net absorption in the regions of strongest UV and \OIII\ emission. We also find that a map of the UV slope $\beta$ in the source plane indicates that the strongest \lya\ emission arises from a very blue region with $\beta \approx -2$, while the brightest UV and \OIII\ emission comes from a redder region with $\beta \approx -1.5$. Source images and the map of the UV slope are shown in Figure \ref{fig:source}.}
\end{enumerate}

\begin{figure}[htbp]
\centerline{\epsfig{angle=00,width=\hsize,file=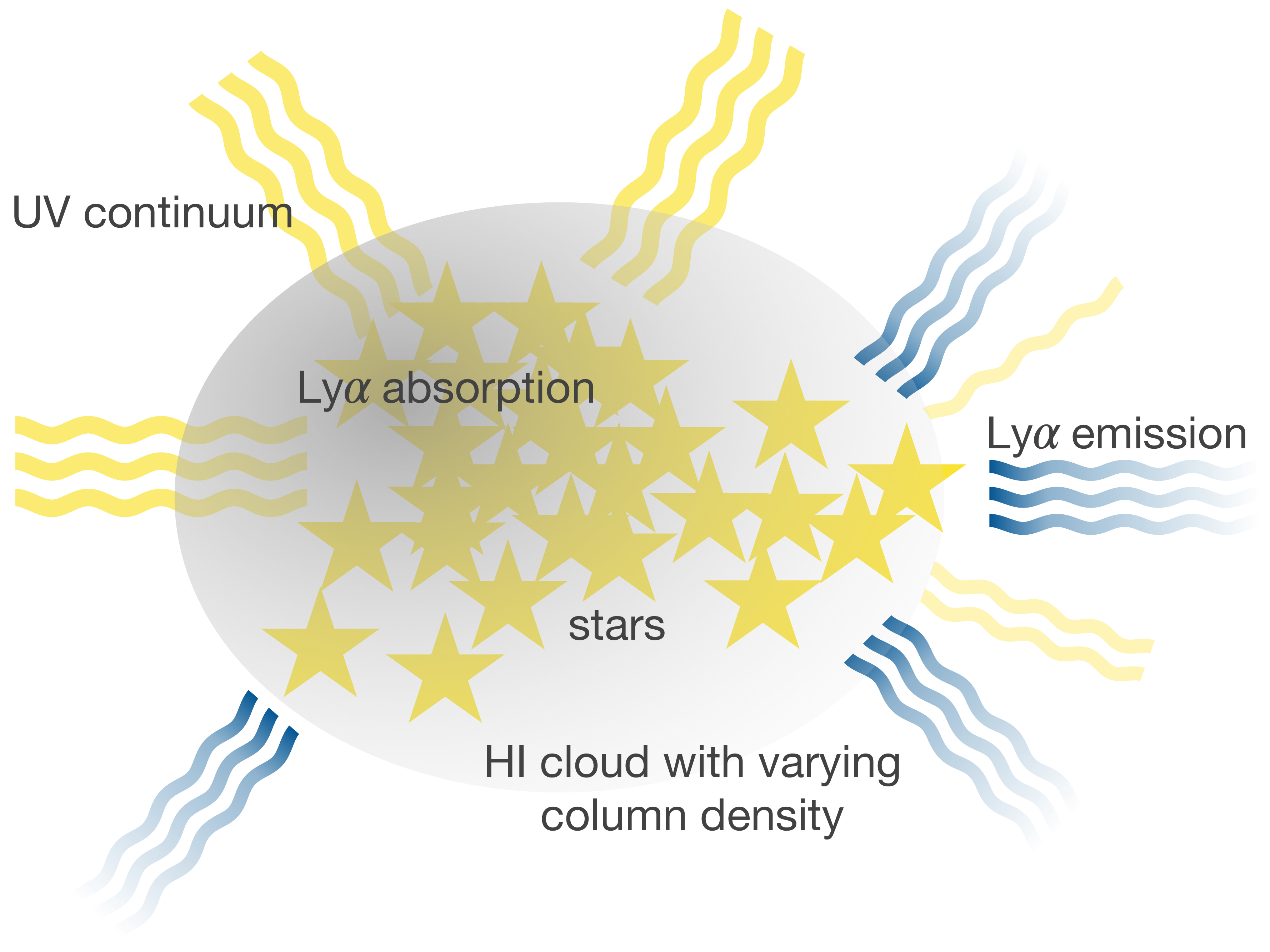}}
\caption{Schematic model of the proposed configuration of stars and neutral hydrogen gas. Intense, compact star formation is covered by a cloud of varying \HI\ column density, resulting in \lya\ absorption in the highest column density regions  and emission in the region of lowest column density. Light from all regions is blended in the spectroscopic slit.}
\label{fig:model}
\end{figure}

We propose a simple schematic model to account for these observations, as shown in Figure \ref{fig:model}. We suggest that several regions of star formation are covered by an \HI\ cloud or clouds of varying column density, and that the highest column density gas is found in the region with the strongest star formation. This region corresponds to the UV and \OIII\ peaks marked in Figure \ref{fig:source}; the high \HI\ column density in this region \edit1{in combination with some dust as indicated by the redder UV slope} then accounts for the net absorption in this portion of the \lya\ image. In the western part of the galaxy the star formation is less intense, as indicated by the weaker UV continuum, but a lower \HI\ column density allows \lya\ emission to escape more easily. This region of lower \HI\ column density also has a blue UV slope $\beta \approx -2$, indicating little dust compared to the regions with stronger star formation. Light from all of these regions is blended together in the spectroscopic slit, resulting in the superposition of strong \lya\ emission and underlying broad absorption.

\edit1{With this model in mind we reassess our previous assumption that the ratio of the continuum in the F343N and F390M filters (the continuum correction factor derived in Equation \ref{eq:ccf}) is constant across the source; if the \lya\ absorption in fact comes primarily from the region of the strongest UV emission, this assumption may not be valid. The quantitative impact of any potential spatial variation in the continuum correction factor is difficult to assess without spatially resolved spectroscopic information, but we do not expect it to have a major impact on our results. If the continuum underlying the \lya\ emission is dominated by the UV-brightest regions, then we should recover the continuum correction factor reasonably well at those locations. However, if the region of peak \lya\ emission actually has little absorption and a higher underlying continuum, the assumption of a constant correction factor would result in an undersubtraction of the continuum in this region, and correcting for this would reduce the emission attributed to \lya. We can gain some insight into the likely impact of this effect from the F343N image shown in Figure \ref{fig:contours}:\ the total F343N image before continuum subtraction demonstrates that the location of the \lya\ peak is not set by the subtraction of the continuum. As shown by the orange contours in the inset panel, the F343N peak nearly coincides with the continuum-subtracted \lya\ peak, and is offset from the peak in the F390M filter. We therefore believe that the spatial offset between the \lya\ and UV emission is robust to uncertainties in the continuum correction factor.}

We now return to the distribution of stars and gas in \sls. While we cannot measure the column density $N_{\rm HI}$ directly, we can gain some insight via comparison with local low metallicity, highly ionized galaxies. The   separation between the blue and red peaks of the \lya\ emission line $\Delta_{\rm peak}$ is sensitive to the column density \citep{d14,vosh15}, and is correlated with the LyC escape fraction in local LyC-emitting galaxies, with most known leakers having $\Delta_{\rm peak} < 400$ \kms\ \citep{iws+18}.  The escape of LyC radiation from these objects indicates at least some channels with $N_{\rm HI} \lesssim 10^{17}$ cm$^{-2}$, the column density at which the gas becomes transparent to the Lyman continuum \citep{zm02}. While it does not necessarily follow that all galaxies with $\Delta_{\rm peak} < 400$ \kms\ have $N_{\rm HI} \lesssim 10^{17}$ cm$^{-2}$, the low peak separation observed for \sls\ ($\Delta_{\rm peak} = 371$ \kms; \citealt{bea+18} and see Figure \ref{fig:filters_spectrum}) does suggest a low column density in the region contributing the \lya\ emission. 

We have proposed that \HI\ gas with higher column density is responsible for the broad absorption feature underlying the \lya\ emission. The continuum S/N is much too low to measure a column density from this absorption feature, but its general shape in the wings of the profile is similar to that of the local low metallicity galaxy SBS 0335$-$052, as we show in Figure \ref{fig:sbs_compare}. \citet{jah+14} measure $\log(N_{\rm HI}/\mbox{cm}^{-2}) = 21.7$ for SBS 0335$-$052, and so we suggest that the spectrum of \sls\ is consistent with a column density $\log(N_{\rm HI}/\mbox{cm}^{-2}) \approx 21.7$ in the regions with net \lya\ absorption.

The superposition of strong \lya\ emission and absorption is not uncommon in extreme, highly ionized galaxies. In a sample of 17 Green Pea galaxies with \OIII$\lambda5007$/\OII$\lambda3727 \geq 6.6$ \edit1{(but somewhat higher metallicities than \sls)}, \citet{mjo+19} find six with both strong, double-peaked \lya\ emission and \lya\ absorption, and determine column densities in the range $19.5 < \log(N_{\rm HI}/\mbox{cm}^{-2}) < 21.4$ by fitting Voigt profiles to the wings of the absorption. Particularly notable is the extreme galaxy J1608, with \OIII/\OII~$=35$ \citep{josd17}; this object shows both absorption with $\log(N_{\rm HI}/\mbox{cm}^{-2}) = 21.4$ and very strong emission with narrow peak separation $\Delta_{\rm peak} =214$ \kms.  The clear implication is that gas with a wide range in column density may exist within the spectroscopic aperture.

Observations indicating a range of \HI\ column densities in a single object are consistent with a model of LyC escape in which ionizing radiation escapes a galaxy through holes with $N_{\rm HI} \lesssim 10^{17}$ cm$^{-2}$ \citep{hbo+11,e15,vosh15,rgd+17,cgs+18,gcs+18,vnc+18,sbs+18,rdc+19}. Although transparent to the Lyman continuum, such gas is optically thick to \lya\ photons, so even the lowest column density gas is likely to produce radiative transfer effects on the \lya\ profile. This model of varying column densities is also supported by measurements of the residual intensities of low ionization interstellar absorption lines, which indicate that a partial covering fraction of neutral gas is common (e.g.,  \citealt{qpss09,hbo+11,rho+15, tssr15,mjo+19}).

\sls\ is consistent with this scenario. Although the low ionization absorption lines are not fully resolved, their depths suggest incomplete coverage of the stellar continuum \citep{bea+18}, while the peak separation of the \lya\ emission line lies in the same range as the values of $\Delta_{\rm peak}$ seen in local LyC-emitters. Higher S/N is needed to confirm the absorption in the \lya\ spectroscopic profile, but the sub-kpc spatial information afforded by the combination of \textit{HST} and gravitational lensing allows us to separate the \lya\ emission from the strongest UV continuum.

It remains to be determined whether or not \sls\ is in fact a LyC-emitter. Although challenging, future space-based observations of the ionizing continuum may address this question. We conclude that \sls\ remains a useful template for the study of reionization-era sources. Its extreme properties mean that it is rare among current samples of gravitationally lensed galaxies, but upcoming facilities such as the Large Synoptic Survey Telescope (LSST) will discover vast numbers of lensed sources \citep{c15}, and extreme objects such as \sls\ will be among them. Observations with the \textit{James Webb Space Telescope} and the 30-m-class telescopes will then provide new constraints on the interplay between stars, radiation and gas in these young galaxies.

\acknowledgements

The authors thank the referee for a thoughtful and constructive report, and John Chisholm, Max Gronke, Matthew Hayes, Jens Melinder, G{\"o}ran {\"O}stlin, and Anne Verhamme for useful discussions. DKE is supported by the US National Science Foundation through the Faculty Early Career Development (CAREER) Program, grant AST-1255591, and by the Wenner-Gren Foundation at Stockholm University. The Cosmic Dawn center is funded by the Danish National Research Foundation.

\facility{{\it HST} (WFC3 UVIS)}
\software{AstroDrizzle \citep{2012AAS...22013515H,2015ASPC..495..281A},  PySynphot \citep{2013ascl.soft03023S}, SExtractor \citep{ba96}}

\bibliographystyle{aasjournal}

\end{document}